\documentclass[doublecol,a4paper,showpacs]{epl2}
\usepackage{graphicx}
\usepackage{amsmath}
\usepackage{amssymb}
\usepackage{enumerate}
\usepackage{subfigure}
\usepackage{tabularx}

\newcommand{\be}{\begin{equation}}
\newcommand{\ee}{\end{equation}}
\newcommand{\ben}{\begin{eqnarray}}
\newcommand{\een}{\end{eqnarray}}
\newcommand{\bes}{\begin{subequations}}
\newcommand{\ees}{\end{subequations}}

\newcommand{\bb}{\bibitem}
\newcommand{\sech}{{\rm sech}}

\begin{document}
\title{Compact Lumps}
\author{D. Bazeia\inst{1}\thanks{Corresponding author; {bazeia@fisica.ufpb.br}} \and M.A. Marques\inst{1}\thanks{{mam.matheus@gmail.com}} \and R. Menezes\inst{2,3}\thanks{{betomenezes@gmail.com}}}
\shortauthor{D.Bazeia \etal}
\institute{
\inst{1}Departamento de F\'\i sica, Universidade Federal da Para\'\i ba, 58051-970 Jo\~ao Pessoa, PB, Brazil\\ 
\inst{2}Departamento de F\'\i sica, Universidade Federal de Campina Grande, 58109-970 Campina Grande, PB, Brazil and \\ 
\inst{3}Departamento de Ci\^encias Exatas, Universidade Federal da Para\'\i ba, 58297-000 Rio Tinto, PB, Brazil.}

\date{\today}
\pacs{11.27.+d}{Extended classical solutions; cosmic strings, domain walls, texture}
\pacs{11.25.-w}{Strings and branes}

\abstract{We study the presence of lumplike solutions in models described by a single real scalar field with standard kinematics in two-dimensional spacetime. The results show several distinct models that support the presence of bell-shaped, lumplike structures which may live in a compact space.}

\maketitle

\section{Introduction}

Defect structures that appear in high energy physics can exhibit a topological or non-topological profile. They have been the issue of several investigations \cite{wilets,vilenkin}, and may also be of interest in many other areas of physics \cite{davidov,murray,walgraef,agrawal}. Usually, topological defects are linearly stable in relativistic field theory, and the non-topological ones are unstable. The instability of the localized structures does not rule them out of physics, because we can enlarge the model and find mechanisms to stabilize the solution \cite{pnevmatikos,xu1,xu2,bnt,bllm,agrawal,haus,frieman,macpherson,coulson,khlopov,coleman,dvali,kusenko,matsuda,bm,sk1,sk2,sk3,R}. An example of this is the case of a fermionic ball \cite{macpherson,bm}, for instance, which can appear when one considers the inclusion of charged fermions, in a way such that the fermions may be entrapped inside the collapsing solution, making it charged and stabilizing the whole structure. Another possibility is to see the scalar field as an axion field \cite{a1,a2,a3}, and this opens interesting new routes \cite{R}. For these reasons, in this work we concentrate mainly on adding another new possibility, of constructing models that support lumplike structures of the compact type,
leaving for the future the use of these solutions to applications in condensed matter and in high energy physics. 

We then focus on the search of non-topological solutions, which appear in models described by a single real scalar field in $(1,1)$ spacetime under the action of nonlinear interactions. They are of general interest, and can be used in soft condensed matter physics  describing, for instance, charge transport in diatomic chains \cite{pnevmatikos,xu1,xu2,bnt,bllm} and bright solitons in fibers \cite{agrawal,haus}, and in high energy physics, for example, as seeds for the formation of structures \cite{frieman,macpherson,coulson,khlopov,bm}, q-balls \cite{coleman,dvali,kusenko,matsuda}, skyrmions \cite{sk1,sk2,sk3} and axions \cite{R}, as particle physics models of inflation \cite{LR} and in braneworld scenarios \cite{B,BJ}. The investigation that we develop is similar to the study implemented in \cite{ave,wes}, but here we focus on another issue, dealing with the presence of lumplike solutions which lives in a compact space.

Defect structures of the compact type firstly appeared in \cite{rosenau}, as solutions of models with nonlinear dispersion and nonlinearity. Since it is not possible to include nonlinear dispersion in relativistic field theory with standard kinematics, it is hard to find compact structures in this case. However, the investigations conducted in \cite{blm} showed the existence of topological compact solutions for specific potentials. Furthermore, in \cite{fktc}, a way to smoothly go from kinks to compactons in models with standard kinematics was developed, although only for topological solutions. 

To close the gap, in this work we focus attention on compact lumps, that is, on non-topological structures of the compact type. The present study is different from the recent work \cite{comp}, where a compact lump is constructed in a Klein-Gordon model. We organize the investigation as follows: In Sec. II we briefly review lumps in the standard scenario and illustrate the results with some explicit examples. We then modify the kinematics of the standard Lagrange density and present compact lumps in Sec. III. We go on and investigate models with standard kinematics, but with specific potentials, showing that it is possible to reach a compact structure for lumps by introducing real and integer parameters in the potential that controls how the scalar field self-interacts. In Sec.~IV we present our conclusions and perspectives.

\section{Lumps}
In order to review models that support non-topological or lumplike solutions, let us consider a Lagrange density with standard kinematics, in the form
\be\label{standardlagrangian}
{\cal L} = \frac12 \partial_\mu\phi \partial^\mu \phi - V(\phi),
\ee
where $\partial_\mu \equiv \partial/\partial x^\mu$, with the Minkowski metric defined by $\eta_{\mu\nu} = \text{diag}(+,-)$, and $V(\phi)$ represents the potential, which we consider to have $V=0$ as a local minimum. Here we consider dimensionless field and coordinates, for simplicity. The equation of motion is 
\be\label{eqmotiontime}
\ddot{\phi}-\phi^{\prime\prime} + V_\phi=0,
\ee
in which the dot and the prime stand for the derivative with respect to $t$ and $x$, respectively, and $V_\phi=dV/d\phi$. If $\phi$ is static field, the above equation simplifies to,
\be\label{eqmotion}
\phi^{\prime\prime} = V_\phi.
\ee
Defining the topological current as $j^\mu = \epsilon^{\mu\nu} \partial_\nu \phi,$ we get the topological charge
$Q = \phi(x\to\infty) - \phi(x\to-\infty).$ Lumps are non-topological solutions, so $Q=0$.  This means that
$\phi(x\to\infty) = \phi(x\to-\infty),$ which can be used as boundary conditions for the equation of motion (\ref{eqmotion}). If a lump exists, we can calculate its energy density to find
\be
\rho(x)=\frac12 {\phi^\prime}^2 + V(\phi).
\ee
In order to see how the static solution behaves under small fluctuations, we study linear stability by taking $\phi(x,t) = \phi(x) + \sum_i \eta_i(x) \cos(\omega_i t)$. We use this into Eq.(\ref{eqmotiontime}) and expand it up to first order in $\eta$ to get a
Schr\"odinger-like equation
\be\label{sta}
-\eta_i^{\prime\prime} + U(x) \eta_i = \omega_i^2 \eta_i,
\ee
with
\be
U(x) = \left. V_{\phi\phi} \right|_{\phi=\phi(x)},
\ee
being the stability potential, which asymptotically tends to the mass of the scalar field at the considered minimum. The zero mode $\eta_0(x)$ can be found to be
\be
\eta_0(x) = \phi^\prime(x).
\ee
In Eq.~(\ref{sta}), the solution is unstable if at least one $\omega_i^2$ is negative. In the case of lumps, the zero mode usually has a node, meaning that there is one lower bound state with negative energy. In the following subsections, we show two examples of lumps in the context of standard kinematics. 

\subsection{Inverted $\phi^4$ model}
\label{invphi4}

To illustrate the general situation, let us consider a model that supports lumplike solutions. We take the inverted $\phi^4$ model, which is given by the potential
\be
V(\phi) = \frac12 \phi^2(1-\phi^2).
\ee
This potential has a local minimum at $\phi=0$ and zeroes at $\phi=\pm1$. The squared mass at the minimum $\phi=0$ is given by $m^2=1$. The equation of motion for static solutions in this case is
\be
\phi^{\prime\prime} = \phi(1-2\phi^2),
\ee
whose solution in the interval $\phi\in[0,1]$, obeying the boundary conditions, is given by
\be
\phi(x) = \sech(x).
\ee
The energy density has the form
\be
\rho(x) = \sech^2(x) \tanh^2(x),
\ee
which gives energy $E=2/3$. Finally, the stability potencial is given by 
\be
U(x) = 1-6\,\sech^2(x).
\ee
It supports the zero mode and another bound state, with negative energy. The zero mode $\eta_0(x) = \sech(x)\tanh(x)$ has a node, thus signaling that it is not the lowest bound state. The lump is unstable under small fluctuations.

\subsection{$\phi^3$ model}\label{phi3}

Another model that admits lump like solutions is the $\phi^3$ model, given by the potential
\be
V(\phi)=2\phi^2(1-\phi).
\ee
It has a local minima at $\phi=0$, with mass $m^2=4$, and a zero at $\phi=1$. The equation of motion for static solutions is
\be
\phi^{\prime\prime} = 2\phi(2-3\phi),
\ee
and now the lump is
\be\label{lumpphi3}
\phi(x) = \sech^2(x),
\ee
with energy density
\be
\rho(x) = 4\,\sech^4(x)\tanh^2(x),
\ee
that can be integrated to give the energy $E=4/3$. The stability potential is given by
\be
U(x) = 4 - 12\sech^2(x).
\ee
It supports the zero mode $\eta(x)=\sech^2(x)\tanh(x)$, which has a node, so there is another mode, with a negative bound state. This shows that the $\phi^3$ lump is unstable under small fluctuations.

\section{Compact Lumps}

Let us now investigate the presence of compact lumps in models described by a single real scalar field.
We first deal with a model with generalized kinematics, and then we study other models, with standard kinematics. 

\subsection{Generalized kinematics}

 We now modify the kinematics of the Lagrange density (\ref{standardlagrangian}) and take a specific potential:
\be
{\cal L} = -\frac14 (\partial_\mu \phi \partial^\mu \phi)^2 - 12 \phi^2(1-\phi)^2.
\ee
The equation of motion that gives the static solutions for this model is
\be
{\phi^\prime}^2\phi^{\prime\prime} = 8\phi(2\phi-1)(\phi-1),
\ee
which is solved by
\begin{equation}\label{cl}
\phi(x)=\left\{
\begin{array}{ll}
\cos^2(x) & |x|\leq\pi/2; \\
0 & |x|> \pi/2.
\end{array} \right.
\end{equation}

The above solution obeys the correct boundary conditions. Then, it is a lump. However, it reaches the boundary values at finite values of $x$, so it is a compact lump. Thus, the solution \textit{per se} only exists in a compact space. This solution exhibits a discontinuity in the second derivative and has energy density
\be 
\rho_c(x)=
\begin{cases}
\sin^4(2x),\,\,\,& |x|\leq\pi/2,\\
0,\,\,\,&|x|>\pi/2,\end{cases}
\ee
that can be integrated to give the total energy $E=3\pi/8$.

The above model supports a non-topological compact structure, but it is described with modified kinematics. However, we want to find compact structures in models with standard kinematics, so in the next subsections, we present models that show this feature.

\begin{figure}[t]
\centering
\includegraphics[width=4.6cm]{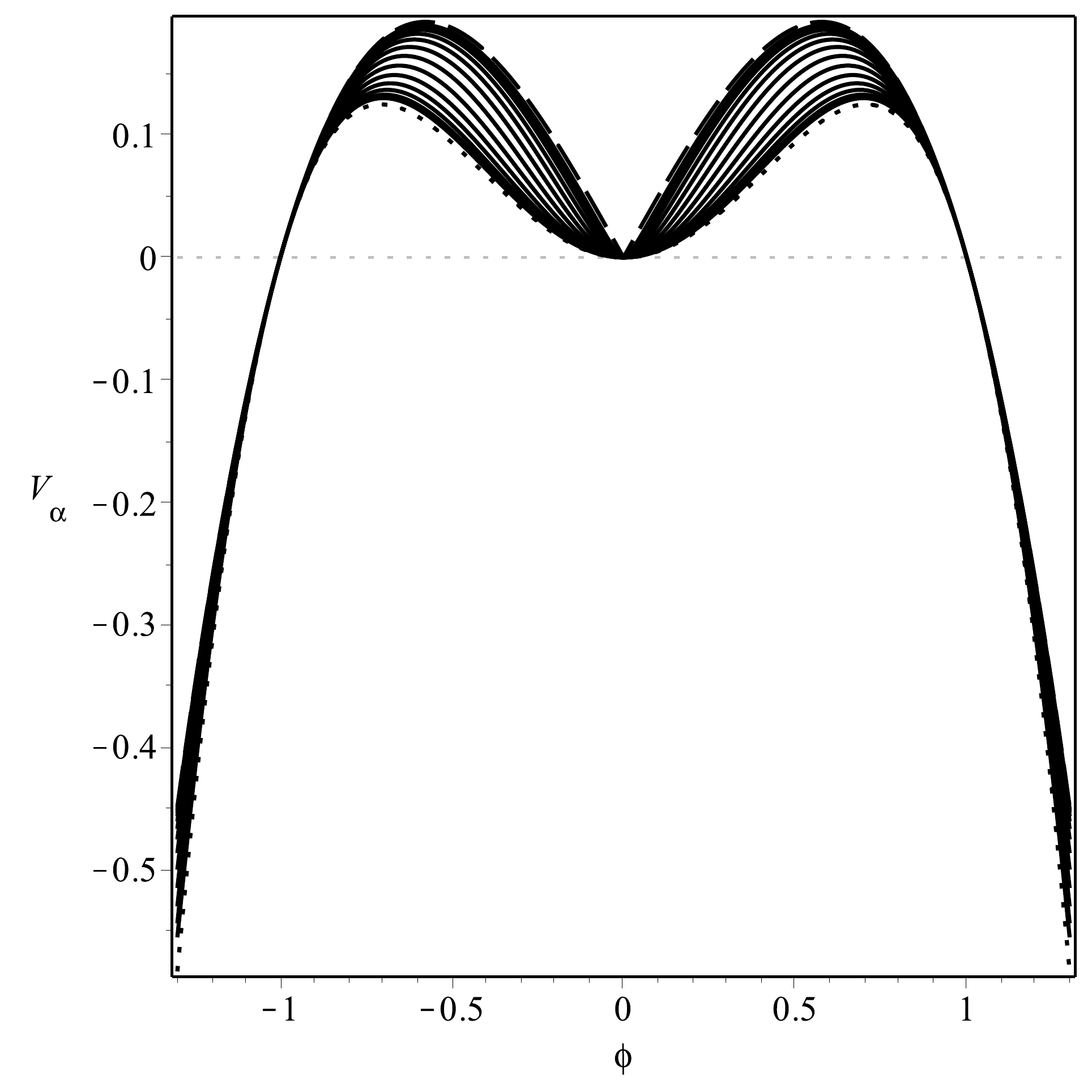}
\caption{The potential (\ref{phi4invalpha}) depicted for $\alpha=0$, and increasing to larger and larger values. The dotted line represents the
$\alpha=0$ case and the dashed line, the $\alpha\to\infty$ case, as in Eq.~(\ref{compactphi4inv}).}
\label{phi4invpot}
\end{figure}

\subsection{Compactifying the inverted $\phi^4$ lump}\label{seccompphi4inv}

We take the Lagrange density (\ref{standardlagrangian}) with the potential
\be\label{phi4invalpha}
V_\alpha(\phi)= \frac{1}{2\alpha}(1-\phi^2)\left(\sqrt{1+\alpha(2+\alpha)\phi^2}-1\right),
\ee
where $\alpha$ is a non-negative real parameter. For any $\alpha$, this potential has a minimum at $\phi=0$, with mass given by $m^2 = 1+\alpha/2$, and zeroes at $\phi=\pm1$. As $\alpha$ increases, the solution tends to compactify. This is similar to the behavior suggested before in \cite{fktc} to describe compact kink. We should note here that a lump has to have the very same behavior asymptotically, and this makes it hard to find asymmetric lumps.

\begin{figure}[t]
\includegraphics[width=4.0cm]{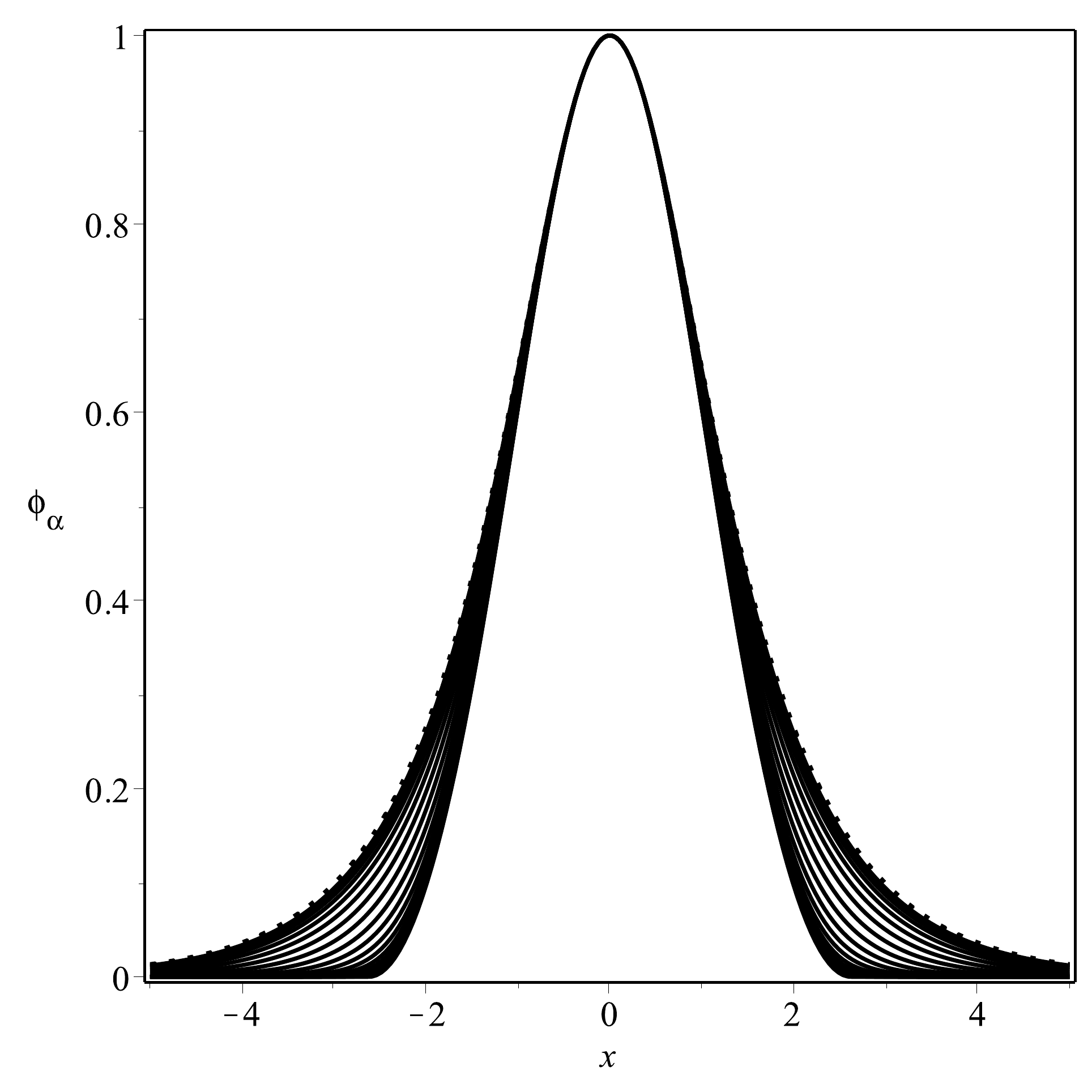}
\includegraphics[width=4.0cm]{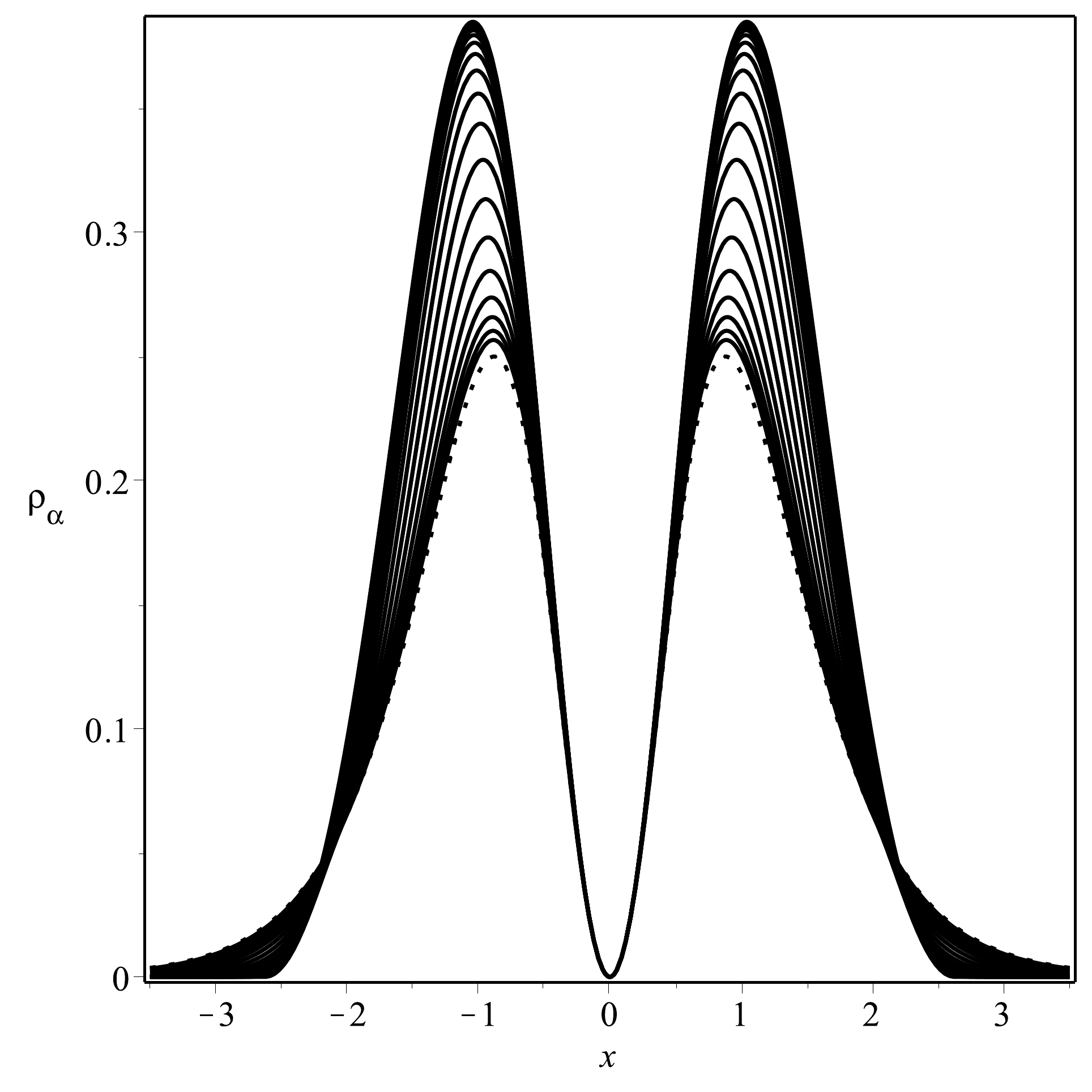}
\caption{The solution of the equation (\ref{eqphi4invalpha}) (left) and its energy density (right) depicted for $\alpha=0$, and increasing to larger and larger values. The dotted line represents the $\alpha=0$ case.}
\label{phi4invsol}
\end{figure}
\begin{figure}[t]
\includegraphics[width=4.0cm]{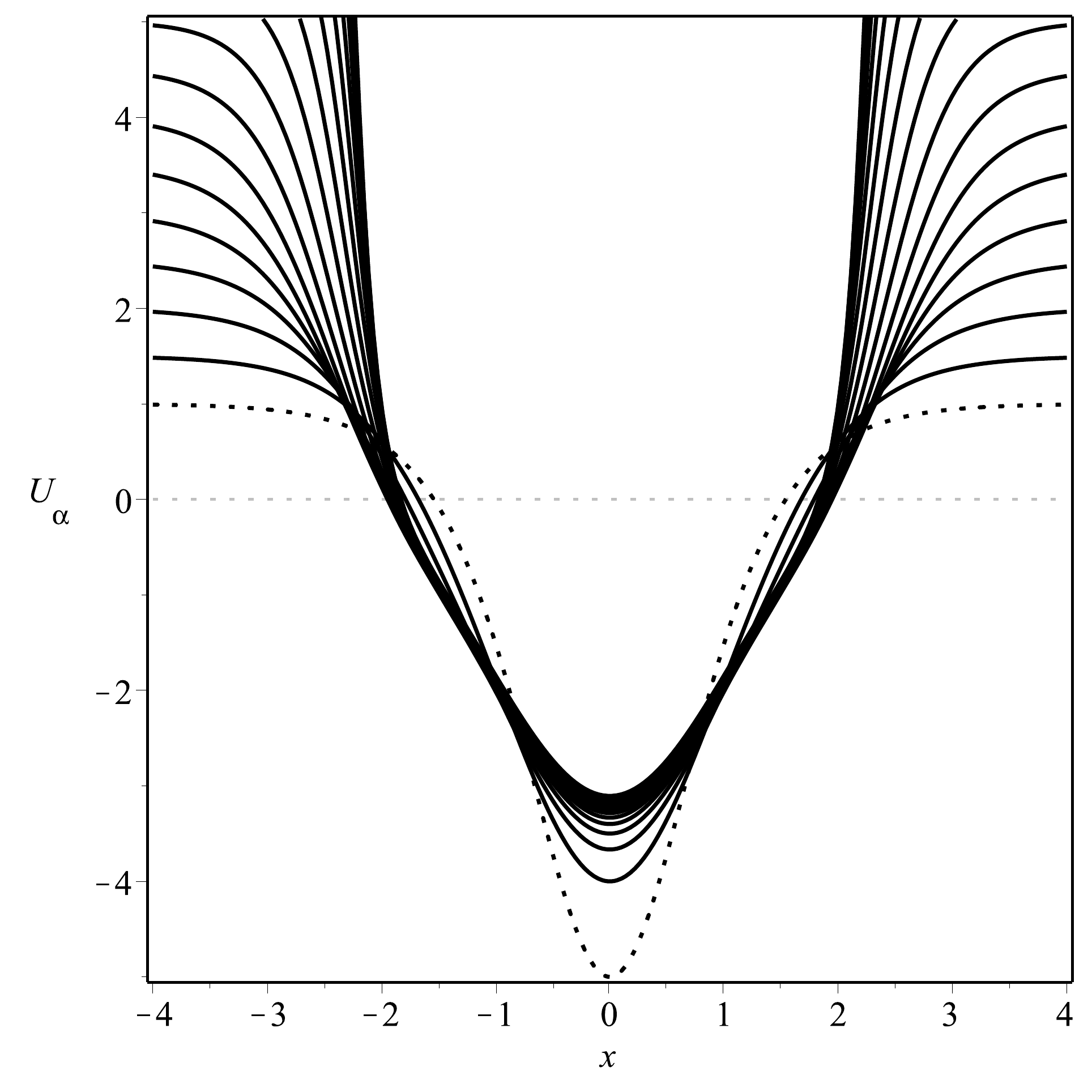}
\includegraphics[width=4.0cm]{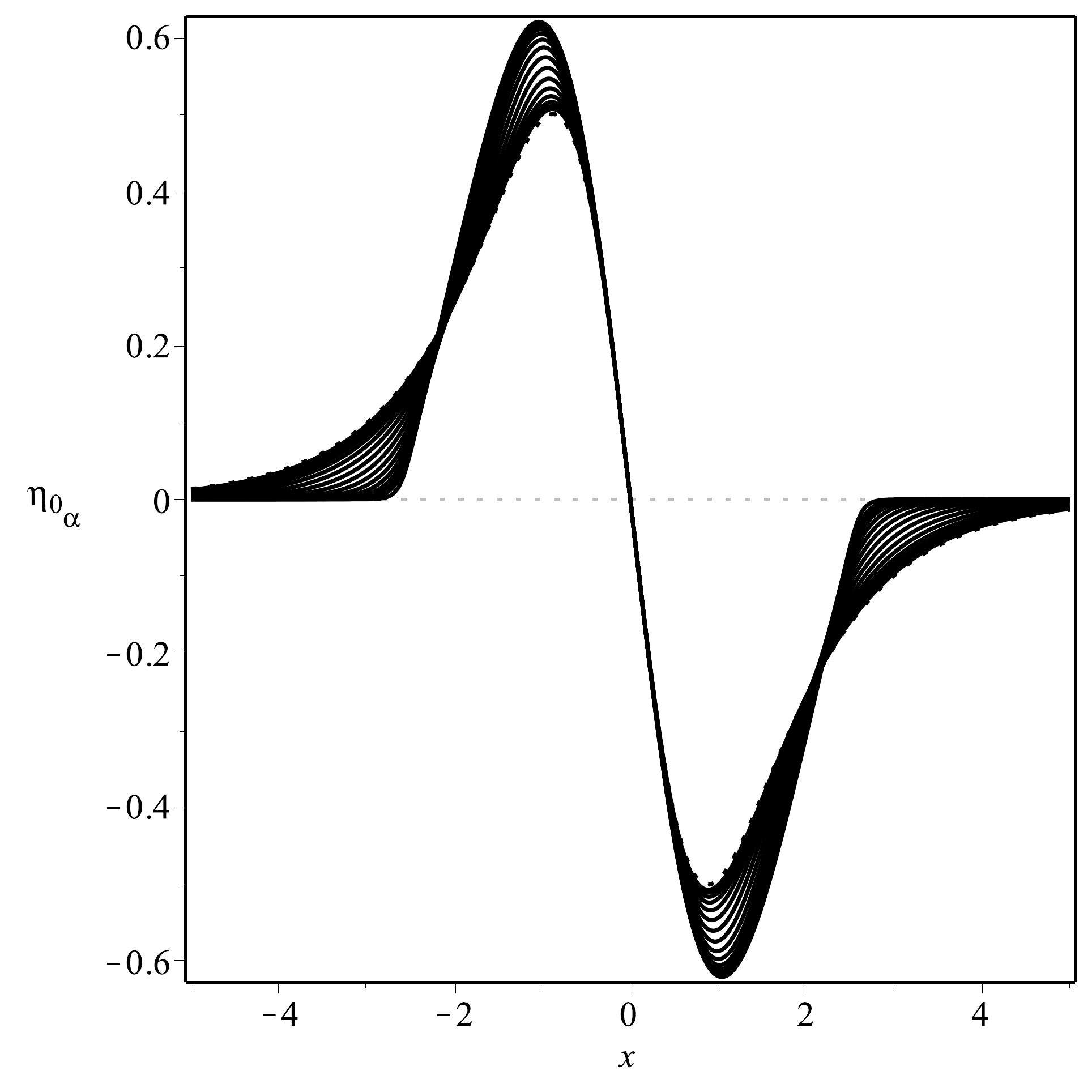}
\caption{The stability potential (left) and the zero modes (right) depicted for $\alpha=0$, and increasing to larger and larger values. The dotted line represents the $\alpha=0$ case.}
\label{phi4invu}
\end{figure}

Let us now look deeper at the potential (\ref{phi4invalpha}). For small $\alpha$, that is, for $\alpha\approx0$ we have
\be
V^s_\alpha(\phi) = \frac12 \phi^2(1-\phi^2) + \frac14\phi^2(1-\phi^2)^2 \alpha + {\cal O}(\alpha^2).
\ee
Then, from the above expression one can see that the limit $\alpha\to0$ gives us the inverted $\phi^4$ model. Also, for large values of $\alpha$, for $\alpha^{-1}\approx0$ we get
\be
V^l_\alpha(\phi)=\frac12|\phi|(1-\phi^2) + \frac12(|\phi| -1)(1-\phi^2)\alpha^{-1} + {\cal O}(\alpha^{-2}).\;\;
\ee
Then, we define
\be\label{compactphi4inv}
V_c(\phi) = \frac12|\phi|(1-\phi^2),
\ee
as the compact limit. In Fig.~\ref{phi4invpot}, we depict the potential (\ref{phi4invalpha}) for $\alpha=0$ and increasing to larger and larger values.

The equation of motion for $\alpha$ arbitrary is
\be\label{eqphi4invalpha}
\phi^{\prime\prime} = \frac{\phi}{2\alpha}\left({\frac{\alpha(2+\alpha)(1-3\phi^2) -2 }{\sqrt {1+\alpha(2+\alpha){\phi}^{2}}}}+2 \right).
\ee
Unfortunately, we have been unable to find analytical solution of the above equation for a general $\alpha$. We only know the solution for the $\alpha=0$ case. The case $\alpha>0$ must be solved numerically. In Fig.~\ref{phi4invsol}, we have depicted the solution and the energy density for several values of $\alpha$, including $\alpha=0$. We note that the energy density has a zero at the origin. By numerical integration, we find that the energy for $\alpha\to\infty$ case is $E_\infty\approx0.9585$. Further, we have plotted the stability potential and the zero modes for this model in Fig.~\ref{phi4invu}. The results show that the model tends to support compact structure in the limit, although we do not know the analytic expression for the solution. We have found numerically that the solution for very large values of $\alpha$ tends to compactify at $x_0\approx2.6221$, and checked that it satisfies the equation of motion at the boundaries.

\begin{figure}[t]
\centering
\includegraphics[width=4.6cm]{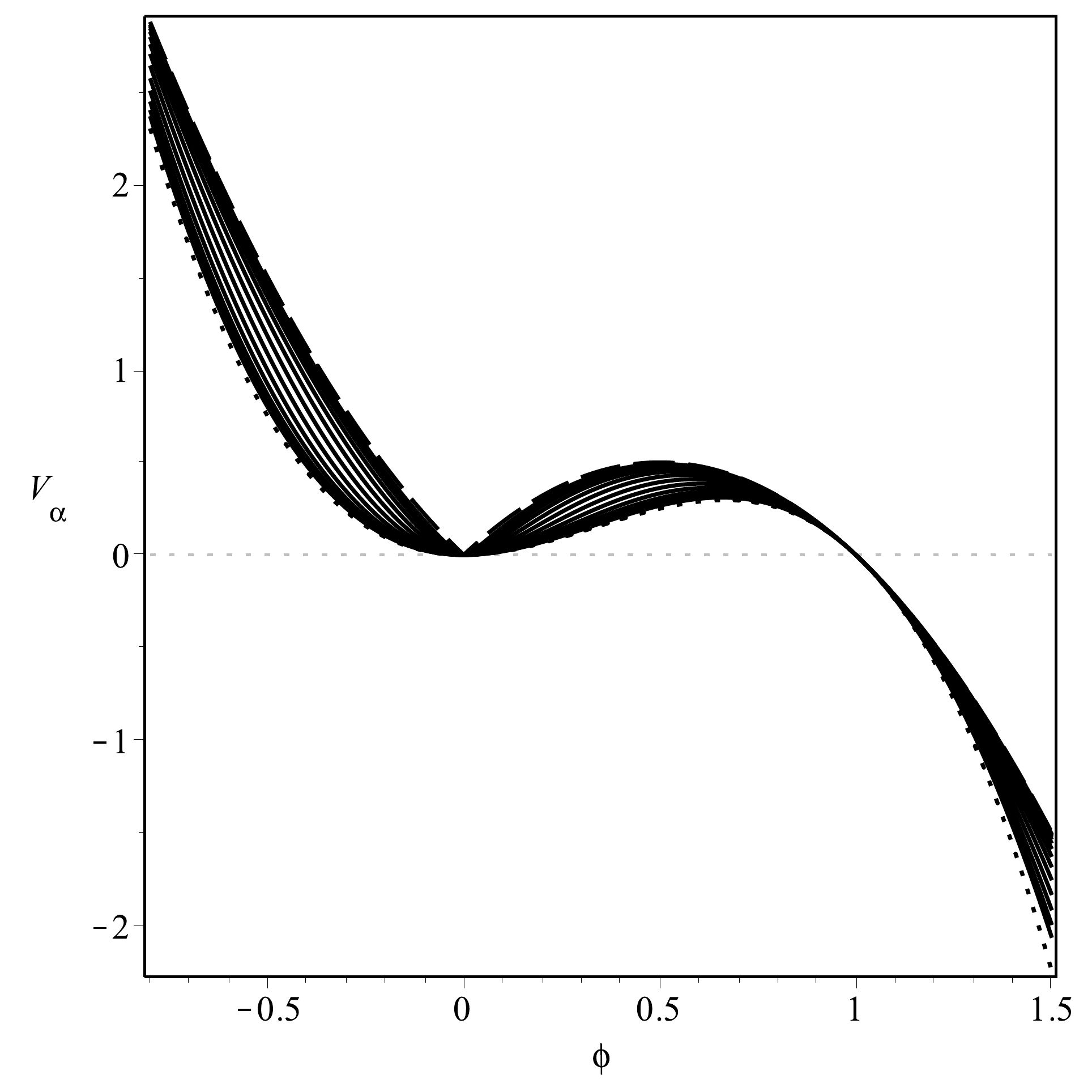}
\caption{The potential (\ref{phi3alpha}) depicted for $\alpha=0$, and increasing to larger and larger values. The dotted line represents the
$\alpha=0$ case and the dashed line, the limit $\alpha\to\infty$, as in Eq.~(\ref{compactphi3}).}
\label{phi3pot}
\end{figure}

\subsection{Compactifying the $\phi^3$ lump}

In the previous subsection, we presented a model that supports compactons, but the analytical expression for the compact solution was found numerically. We now investigate a model that admits an analytical expression for the compact solution. This can be done with the standard Lagrange density and the new potential
\be\label{phi3alpha}
V_\alpha(\phi)= \frac{2}{\alpha}(1-\phi)\left(\sqrt{1+\alpha(2+\alpha)\phi^2}-1\right),
\ee
where, again, $\alpha$ is a non-negative real parameter. For any $\alpha$, this potential has a minimum at $\phi=0$, with mass given by $m^2 = 4+2\alpha$, and a zero at $\phi=1$. Let us see how this potential behaves when $\alpha$ is small, that is, $\alpha \approx 0$:
\be
V^s_\alpha(\phi) = 2\phi^2(1-\phi) + \phi^2(1-\phi)(1-\phi^2)\alpha + {\cal O}(\alpha^2).
\ee
From the above equation, we see that the limit $\alpha\to0$ leads us to the $\phi^3$ model already investigated. On the other hand, the asymptotic behavior $(\alpha^{-1} \approx 0)$ of the potential is
\be
V^l_\alpha(\phi)= 2|\phi|(1-\phi) + 2(1-\phi)(|\phi|-1)\alpha^{-1} + {\cal O}(\alpha^{-2}),
\ee
so we take the potential
\be\label{compactphi3}
V_c(\phi) = 2|\phi|(1-\phi),
\ee
to define the compact limit. The potential (\ref{phi3alpha}) is depicted in Fig.~\ref{phi3pot} for several values of $\alpha$.

\begin{figure}[t]
\includegraphics[width=4.0cm]{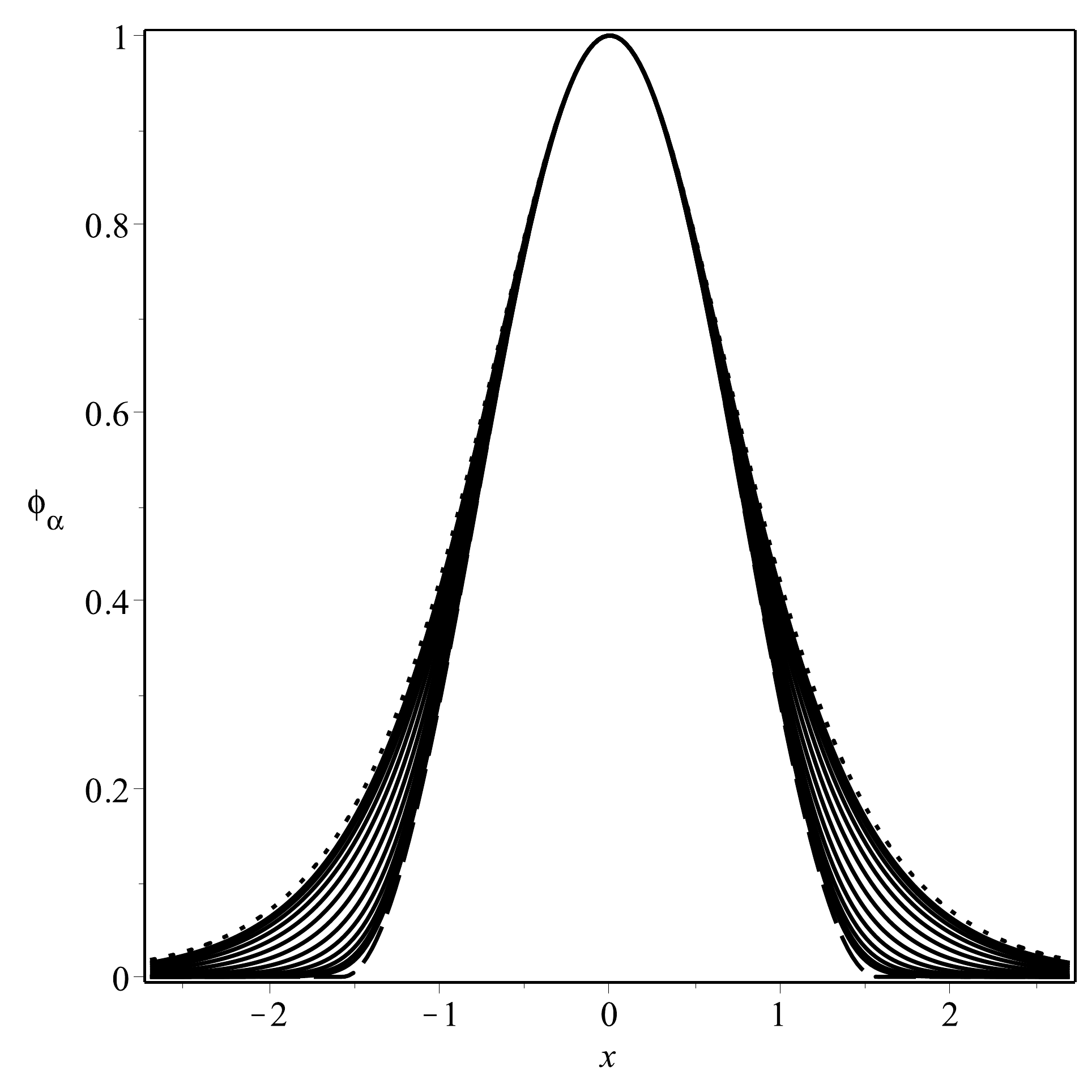}
\includegraphics[width=4.0cm]{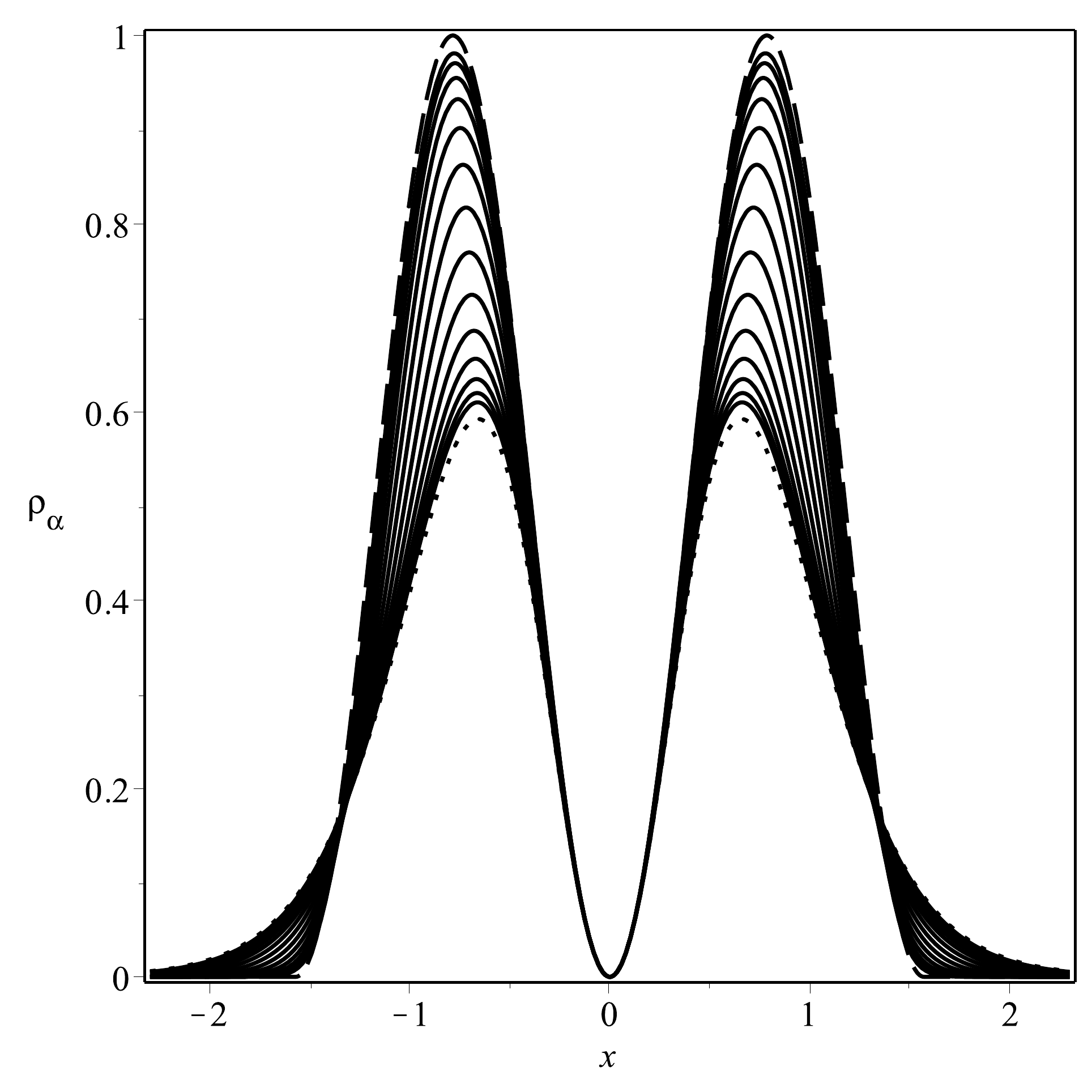}
\caption{The solution of the equation (\ref{eqphi3alpha}) (left) and its energy density (right) depicted for $\alpha=0$, and increasing to larger and larger values. The dotted and dashed lines represent the $\alpha=0$ and $\alpha\to\infty$ cases, respectively.}
\label{phi3sol}
\end{figure}

The equation of motion for $\alpha$ arbitrary is
\be\label{eqphi3alpha}
\phi^{\prime\prime} = \frac{2}{\alpha}\left({\frac{\alpha(2+\alpha)(1-2\phi)\phi -1 }{\sqrt {1+\alpha(2+\alpha){\phi}^{2}}}}+1\right).
\ee
For general $\alpha$, this equation was solved numerically and its solutions, as well as its energy density, are plotted in Fig.~\ref{phi3sol}. Also,  we have checked that the numerical solutions satisfy the equation of motion at the boundaries.

\begin{figure}[t]
\includegraphics[width=4.0cm]{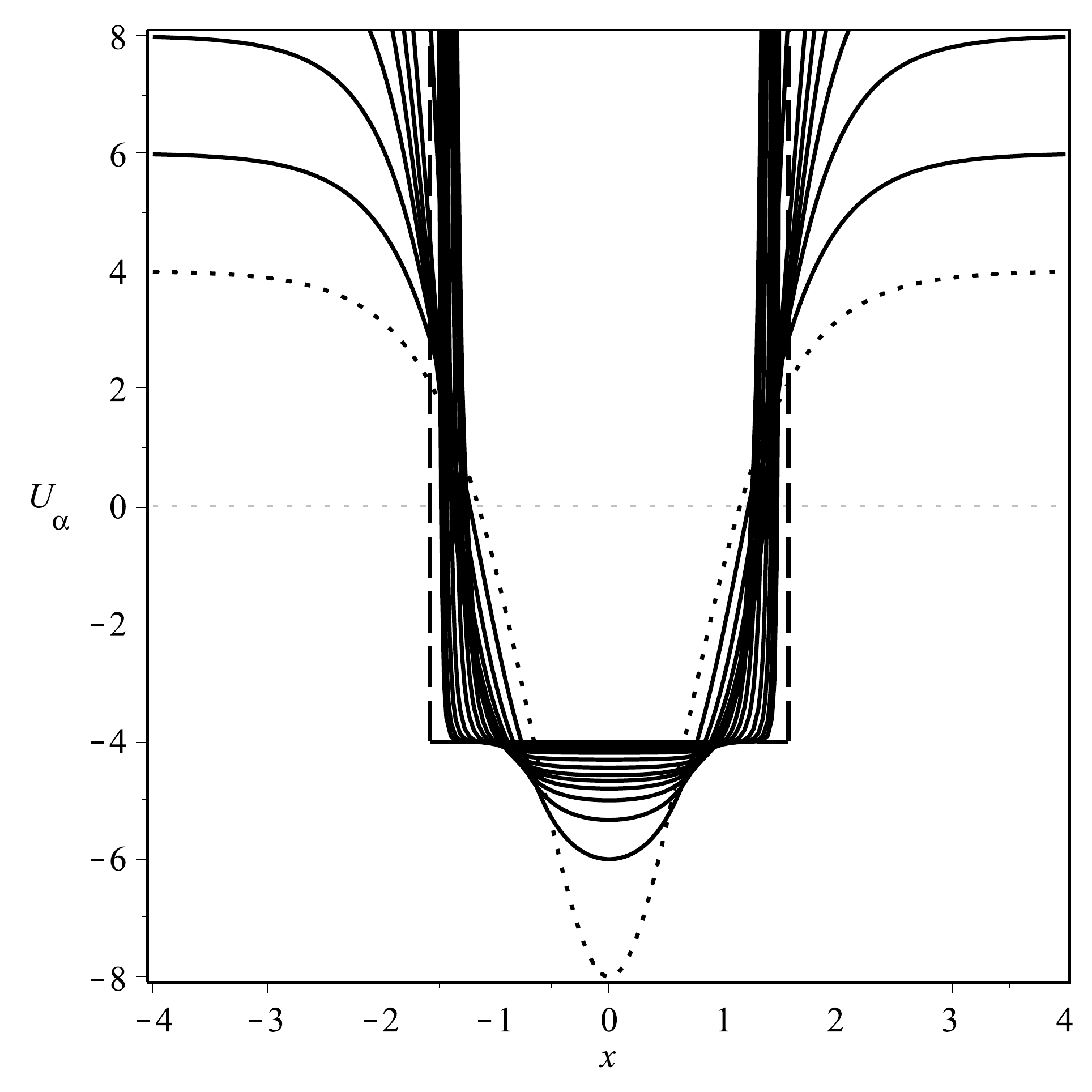}
\includegraphics[width=4.0cm]{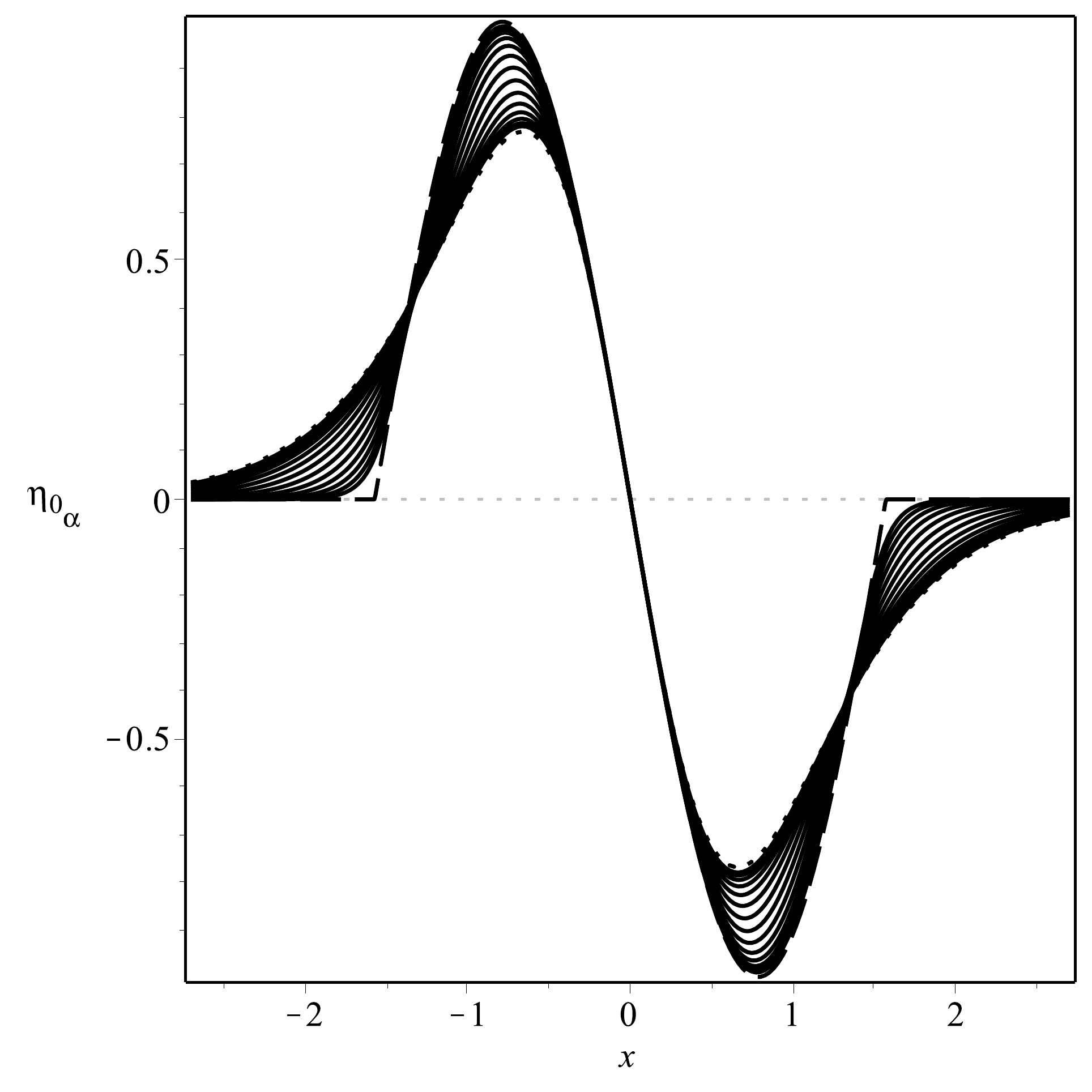}
\caption{The stability potential (left) and the zero modes (right) depicted for $\alpha=0$, and increasing to larger and larger values. The dotted and dashed lines represent the $\alpha=0$ and the $\alpha\to\infty$ case, respectively.}
\label{phi3u}
\end{figure}

However, one can calculate the equation of motion for the case $\alpha\to\infty$, represented by the potential (\ref{compactphi3}) to get:
\be\label{eqphi3compact}
\phi^{\prime\prime} = \frac{2\phi}{|\phi|} (1 - 2\phi).
\ee
This equation admits exactly the solution (\ref{cl}), which is a compact lump. We note that for the solution \eqref{cl}, both sides of \eqref{eqphi3compact} go to $2$, when $x\to\pm\pi/2$. This shows that we have compactified the solution (\ref{lumpphi3}) in a scenario with standard kinematics. As we have the expression for the compact solution, we can calculate its energy density:
\be 
\rho_c(x)=
\begin{cases}
\sin^2(2x),\,\,\,& |x|\leq\pi/2,\\
0,\,\,\,&|x|>\pi/2,\end{cases}
\ee
that can be integrated all over the space to give the energy $E_c=\pi/2$. Furthermore, the stability potential is
\be
U_c(x)=
\begin{cases}
-4,\,\,\,& |x|\leq\pi/2,\\
\infty,\,\,\,&|x|>\pi/2,\end{cases}
\ee
and the zero mode has the form
\be
{\eta_c}_0(x)=
\begin{cases}
-\sin(2x),\,\,\,& |x|\leq\pi/2,\\
0,\,\,\,&|x|>\pi/2.\end{cases}
\ee
In Fig.~\ref{phi3u}, we have depicted the stability potential and the zero modes for several values of $\alpha$, including the cases $\alpha=0$ and $\alpha\to\infty$.

\begin{figure}[htb!]
\centering
\includegraphics[width=4.6cm]{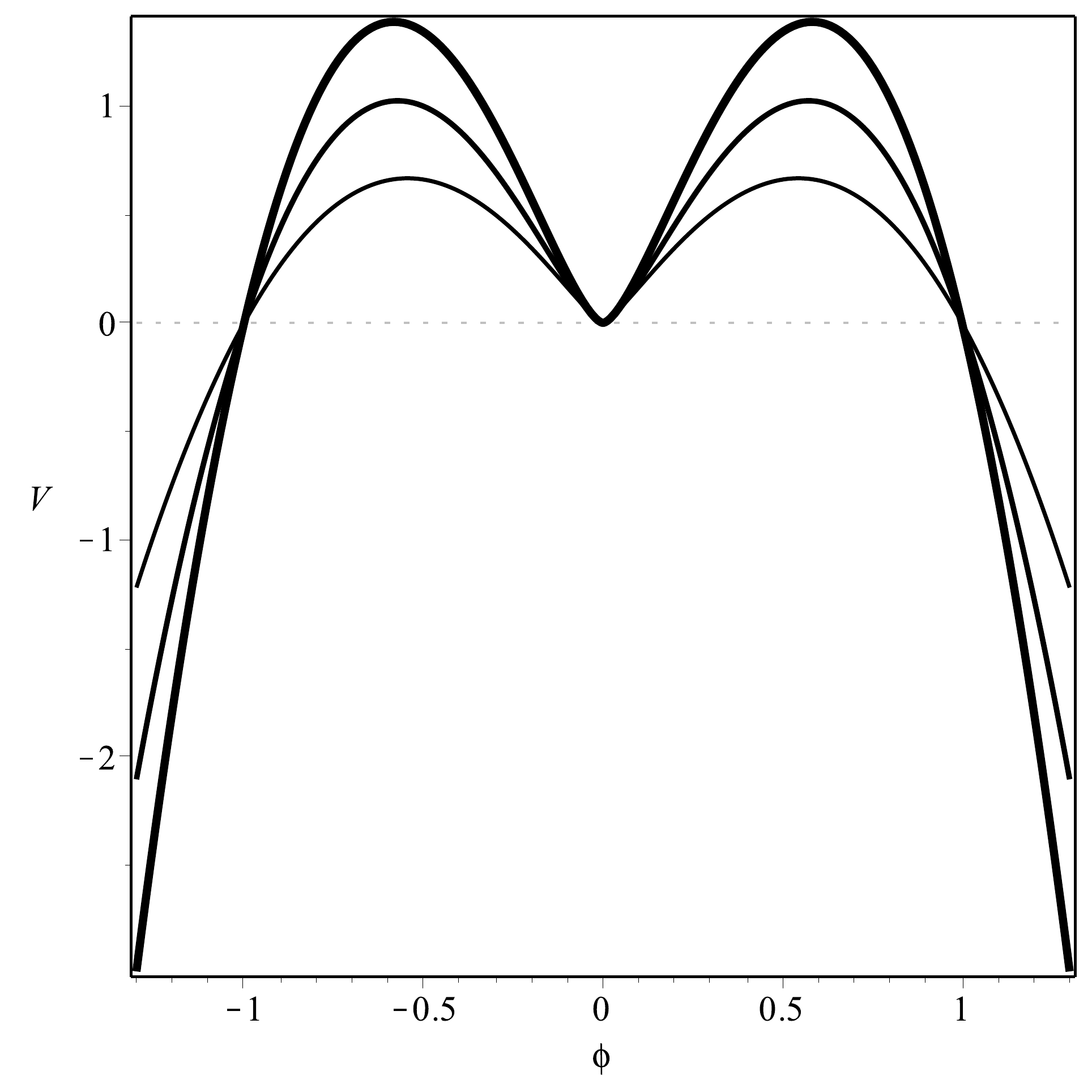}
\caption{The potential (\ref{qmodelpot}) depicted for $q=3,5$ and $7$. The thickness of the lines increases with $q$.}
\label{qpot}
\end{figure}

\subsection{Analytical model}

We now introduce another model, described by the potential
\be\label{qmodelpot}
V(\phi)= \frac12{q^2}\left(\phi^{-2/q}-1\right)\phi^2,
\ee
where $q=3,5,7,\cdots$. This potential has a minimum at $\phi=0$ and two zeroes at $\phi=\pm1$. It is depicted in Fig.~\ref{qpot}. 
It is similar to the case considered in \cite{bmm}, used to give rise to new kinklike structures of the 2-kink type.
It is also similar to the case considered in \cite{comp}, used to build charged compact structures in a relativistic Klein-Gordon model.
It is important here because it can be solved analytically for $q=3,5,7,\cdots,$ arbitrary. 

\begin{figure}[t]
\includegraphics[width=4.0cm]{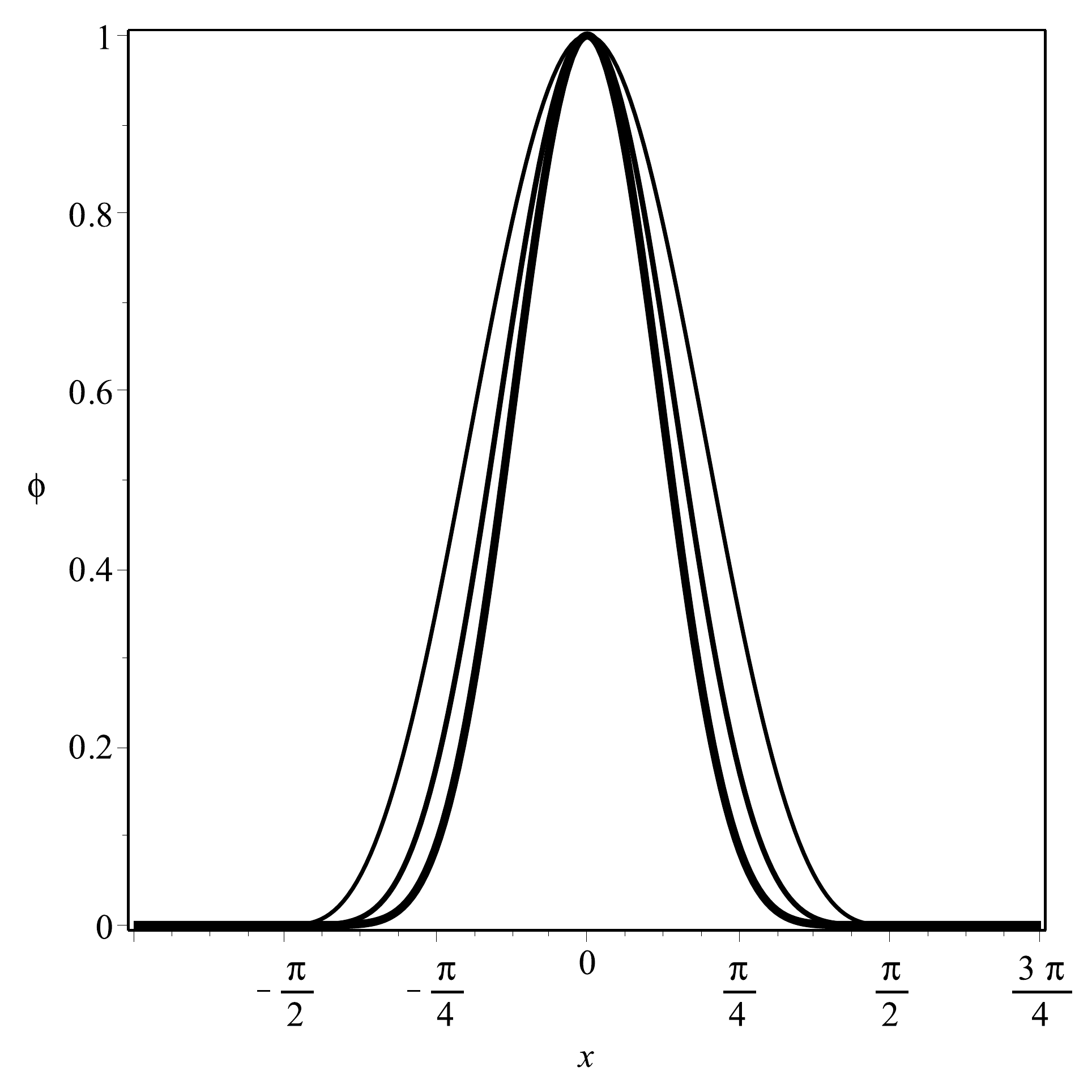}
\includegraphics[width=4.0cm]{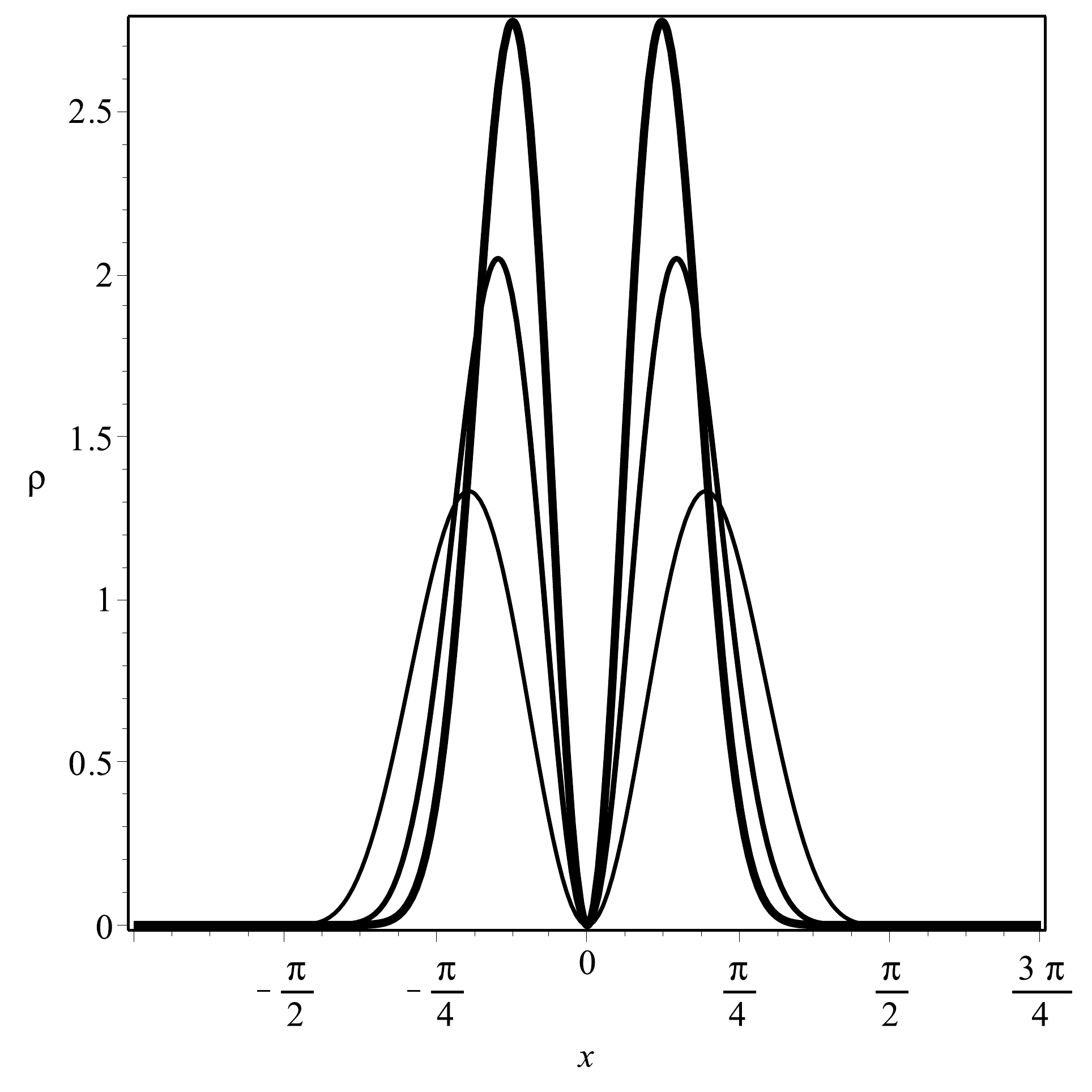}
\caption{The solution (\ref{qmodelsol}) (left) and its energy density (\ref{qmodelrho}) (right) depicted for $q=3,5$ and $7$. The thickness of the lines increases with $q$.}
\label{qsol}
\end{figure}

The equation of motion for static solutions is
\be
\phi^{\prime\prime} = -q\phi\left(q+(1-q)\phi^{-2/q}\right).
\ee
It admits the solution
\be\label{qmodelsol}
\phi_q(x)=\begin{cases}
\cos^q(x), & |x|\leq \frac{\pi}{2}, \\
0, & |x|>\frac{\pi}{2},
\end{cases}
\ee
which is a compact lump. The energy density is
\be\label{qmodelrho}
\rho_q(x)=\begin{cases}
\sin^2(x)\cos^{2q-2}(x), & |x|\leq \frac{\pi}{2}, \\
0, & |x|>\frac{\pi}{2},
\end{cases}
\ee
that can be integrated to give energy
\be
E_q=\frac{\sqrt{\pi}}{2} \frac{\Gamma(q-\frac12)}{\Gamma(q+1)},
\ee
where $\Gamma(z)$ is the Gamma Function. In Fig.~\ref{qsol}, we have depicted the solution and the energy density for some values of $p$.

\begin{figure}[t]
\includegraphics[width=4.0cm]{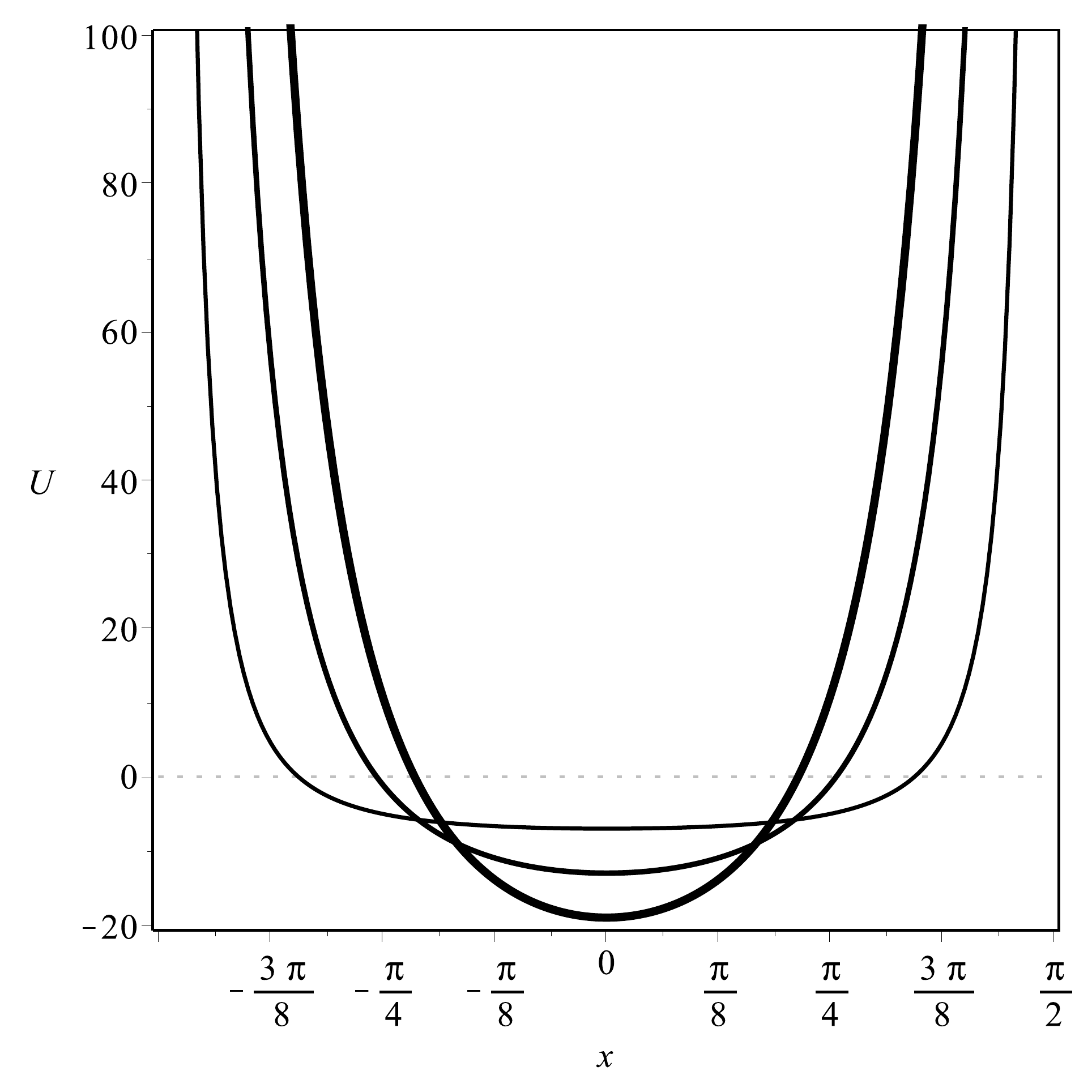}
\includegraphics[width=4.0cm]{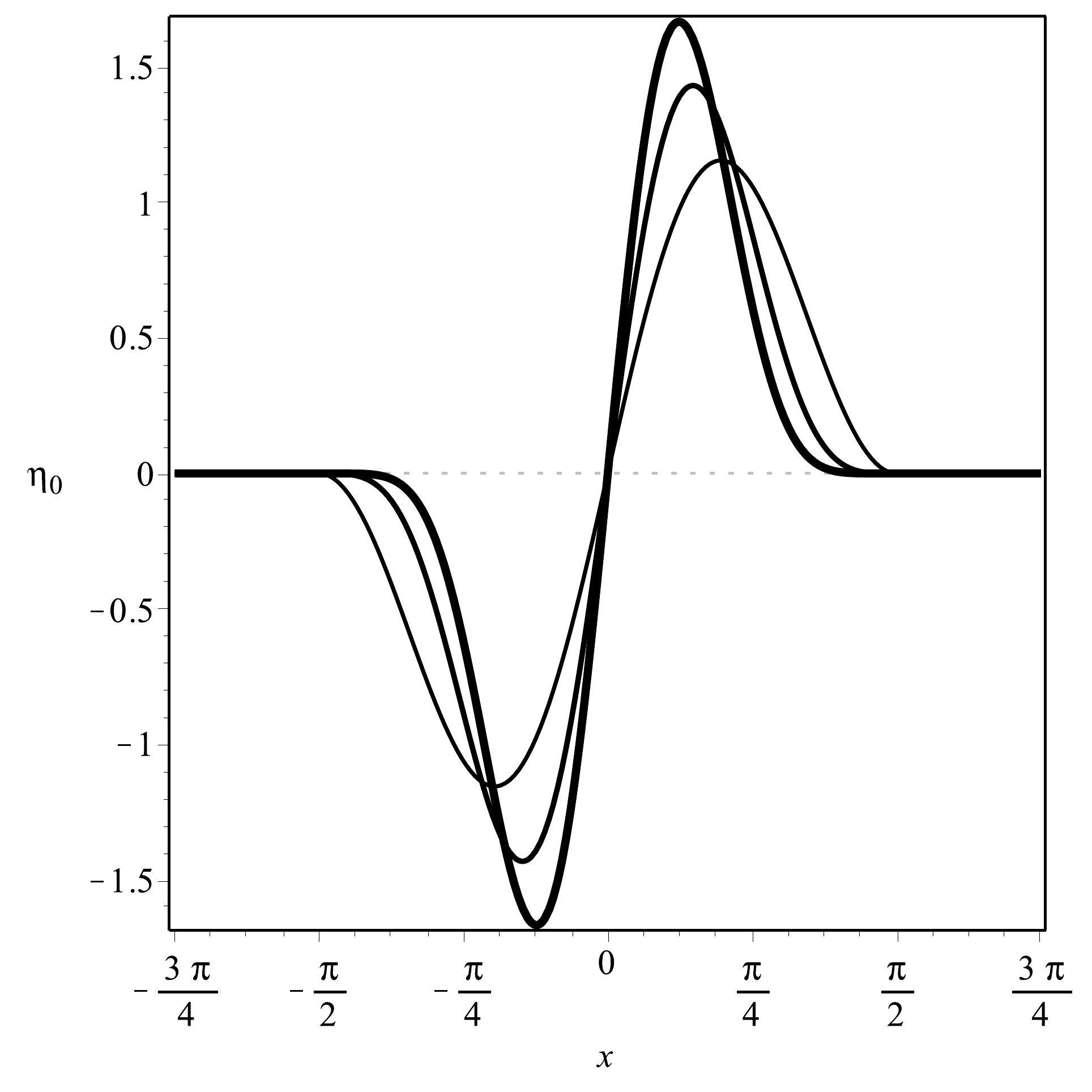}
\caption{The stability potential (\ref{qmodelU}) (left) and its zero mode (\ref{qmodeleta}) (right) depicted for $q=3,5$ and $7$. The thickness of the lines increases with $q$.}
\label{qU}
\end{figure}

In this case, the stability potential is given by
\be\label{qmodelU}
U_q(x)=\begin{cases}
(q-1)(q-2)\sec^2(x)-q^2, & |x|\leq \frac{\pi}{2}, \\
\infty, & |x|>\frac{\pi}{2},
\end{cases}
\ee
and the zero mode is
\be\label{qmodeleta}
\eta_q(x)=\begin{cases}
q\sin(x)\cos^{q-1}(x), & |x|\leq \frac{\pi}{2}, \\
0, & |x|>\frac{\pi}{2}.
\end{cases}
\ee

In Fig.~\ref{qU} we can see how the stability potential and zero mode behave. The zero mode presents a node, which is an evidence of the instability of the lump like solution, as it is for lumps in general.

This model is different from the two others, since it is solved analytically, and supports compact solution for every $q=3,5,7,\cdots.$  In the two previous cases, we get to a compact solution asymptotically, as we make the real parameter to increase to larger and larger values. The presence of analytical solutions helps us to explore the stability potential (\ref{qmodelU}) analytically. It is the P\"oschl-Teller potential \cite{pt,flugge}, and the eigenvalues are given by
\be  
\omega_n^2=(n-1)(2q+n-1),
\ee
for $n=0,1,2,\dots$. We note that $\omega_0^2=1-2q$, which is clearly negative since $q=3,5,7,\ldots$; thus, the lumplike solutions (\ref{qmodelsol}) are unstable. The eigenfunctions corresponding to the above eigenvalues can be obtained analytically; they are
\ben\label{modes}
\eta_{n,q}(x)&\!\!=\!\!&C_{n,q}\;{}_2 F_1\!\left(-n,2q\!+\!n-2,q\!-\!\frac12,\sin^2\left(\frac{x}{2}\!+\!\frac{\pi}{4}\right)\!\right) \nonumber\\
&&\!\! \times\cos^{q-1}(x),
\een
where ${}_2 F_1\left(a,b,c,z\right)$ is Hypergeometric function, $C_{n,q}$ stands for the normalization constant, and $x$ is rectricted to the compact interval $[-\pi/2,\pi/2]$. We note from Eq.~(\ref{modes}) that $\eta_{1,q}$ equals $\eta_q$ given by Eq.~(\ref{qmodeleta}), as expected. 

\section{Ending comments}

We studied the existence of compact lumps in relativistic field theory. After reviewing the basic facts about lumps, we investigated a model with modified kinematics, showing that it gives rise to compact lumplike structures. We then showed how to make a lump compact, in models with standard kinematics. We studied three distinct models, two described by a real parameter, one inspired on the inverted $\phi^4$ model, which is solved numerically, and the other which is inspired on the $\phi^3$ model, also solved numerically, but with the compact limit allowing for a analytic compact solution, and the third one, described by a different parameter, $q$, which can be odd integer. The third model is particularly interesting, since it can be solved analytically and presents compact solutions for every (odd integer) value of the parameter $q$. 

The models that we investigated describe lumplike structures, so all the solutions are linearly unstable, with the corresponding zero modes having a node at the origin. However, we can stabilize them with the addition of charged fermions, for instance, or changing the model to describe complex field and searching for q-balls and other localized structures in one, two or three spatial dimensions. Moreover, we can consider lumplike excitations as axions in the dark sector \cite{R} and as tachyonic branes in curved spacetime with a warped geometry \cite{B,BJ}. The fact that we found compact objects naturally induces the compact behavior, without the need to make the spatial dimension compact. These and other issues motivate us to go further and extend the above models, searching for such compact structures in other scenarios.

\acknowledgements{We would like to thank the Brazilian agency CNPq for partial financial support. DB thanks support from fundings 455931/2014-3 and 06614/2014-6, MAM thanks support from funding 140735/2015-1, and RM thanks support from fundings 508177/2010-3 and 455619/2014-0.}


\begin{thebibliography}{99}
\bibitem{wilets} L. Wilets, Non topological solitons, World Scientific (1989).
\bibitem{vilenkin} A. Vilenkin and E.P.S. Shellard, Cosmic Strings and Other Topological Defects, Cambridge University Press (2007).
\bibitem{davidov} A.S. Davidov, Solitons in Molecular Systems, Kluwer (1981).
\bibitem{murray} J.D. Murray, Mathematical Biology, Springer (1989).
\bibitem{walgraef} D. Walgraef, Spatio-Temporal Pattern Formation, Springer (1997).
\bibitem{a1}R.D. Peccei and H.R. Quinn, Phys. Rev. Lett. {\bf38} 1440 (1977).
\bibitem{a2}S. Weinberg, Phys. Rev. Lett. {\bf40}, 223 (1978).
\bibitem{a3}F. Wilczek, Phys. Rev. Lett. {\bf40}, 279 (1978).
\bibitem{agrawal} G.P. Agrawal, Nonlinear Fiber Optics, Academic (1995).
\bibitem{pnevmatikos} S. Pnevmatikos, Phys. Rev. Lett. 60, 1534 (1988).
\bibitem{xu1}J.-Z. Xu and J.-N. Huang, Phys. Lett. A 197, 127 (1995).
\bibitem{xu2}J.-Z. Xu and B. Zhou, Phys. Lett. A 210, 307 (1996).
\bibitem{bnt}D. Bazeia, J.R. Nascimento, and D. Toledo, Phys. Lett. A 228, 357 (1997).
\bibitem{bllm}D. Bazeia, V.B.P. Leite, B.H.B. Lima, and F. Moraes, Chem. Phys. Lett. 340, 205 (2001).
\bibitem{haus}H.A. Haus and W.S. Wong, Rev. Mod. Phys. 68, 423 (1996).
\bibitem{frieman}J.A. Frieman, G.B. Gelmini, M. Gleiser, and E.W. Kolb, Phys. Rev. Lett. 60, 2101 (1988).
\bibitem{macpherson}A.L. MacPherson and B.A. Campbell, Phys. Lett. B 347, 205 (1995).
\bibitem{coulson}D. Coulson, Z. Lalak, and B. Ovrut, Phys. Rev. D 53, 4237 (1996).
\bb{bm}J.R. Morris and D. Bazeia, Phys. Rev. D {54}, 5217 (1996).
\bibitem{khlopov}M.Y. Khlopov, Cosmoparticle Physics, World Scientific (1999).
\bibitem{coleman}S. Coleman, Nucl. Phys. B 262, 263 (1985).
\bibitem{dvali}G. Dvali, A. Kusenko, and M. Shaposhnikov, Phys. Lett. B 417, 99 (1998).
\bibitem{kusenko}A. Kusenko and M. Shaposhnikov, Phys. Lett. B 418, 46 (1998).
\bibitem{matsuda}T. Matsuda, Phys. Rev. D 68, 127 302 (2003).
\bb{sk1}J. Ashcroft, M. Haberichter, and S. Krusch, Phys. Rev. D {91}, 105032 (2015). 
\bb{sk2}C. Adam, C. Naya, J. Sanchez-Guillen, and A. Wereszczynski, Phys. Rev. D 86, 085001 (2012).
\bb{sk3}J.M. Speight, J. Phys. A 43, 405201 (2010). 
\bibitem{R}A. Ringwald, Phys. Dark Univ. {1}, 116 (2012).
\bb{LR}D.H. Lyth and A. Riotto, Phys Rept. 314, 1 (1999).
\bb{B}F.A. Brito, JHEP {0508}, 036 (2005).
\bb{BJ} F.A. Brito and H.S. Jesuino, JHEP 1007, 031 (2010).
\bb{ave}A.T. Avelar, D. Bazeia, L. Losano, and R. Menezes, Eur. Phys. J. C 55, 133 (2008).
\bb{wes}A.T. Avelar, D. Bazeia, W.B. Cardoso, and L. Losano, Phys. Lett. A 374, 222 (2009).
\bibitem{rosenau}P. Rosenau and J.M. Hyman, Phys. Rev. Lett. 70, 564 (1993).
\bibitem{blm} D. Bazeia, L. Losano, and R. Menezes, Phys. Lett. B 731, 293 (2014);
D. Bazeia, L. Losano, M. A. Marques, and R. Menezes, EPL 107, 61001 (2014).
\bibitem{fktc} D. Bazeia, L. Losano, M. A. Marques, and R. Menezes, Phys. Lett. B 736, 515 (2014).
\bb{comp}P. Rosenau and E. Kashdan, Phys. Rev. Lett. 104, 034101 (2010).
\bb{bmm}D. Bazeia, J. Menezes, and R. Menezes, Phys. Rev. Lett. 91, 241601 (2003).
\bb{pt} G. P\"{o}schl and E. Teller, Z. Physik 83, 143 (1933).
\bb{flugge}S. Fl\"ugge, Practical Quantum Mechanics, Springer (1998).
\end{thebibliography}
\end{document}